\documentclass{llncs}
\usepackage{amsmath,amssymb}
\usepackage{url}
\usepackage{epsfig}
\usepackage[latin1]{inputenc}
\usepackage[mathscr]{euscript}
\usepackage{subfigure}
\usepackage{wrapfig}

\input xy \xyoption{all}
\newcommand{\comment}[1]{{}}

\newcommand{\calx}{\mathcal{X}}
\newcommand{\caly}{\mathcal{Y}}
\newcommand{\calz}{\mathcal{Z}}

\newcommand{\calu}{\mathcal{U}}

\usepackage{color}

\begin{document}

\title{On the relation between Differential Privacy and Quantitative Information Flow\thanks{This work has been partially supported by the project ANR-09-BLAN-0169-01  PANDA  and by  the INRIA DRI  Equipe Associ\'ee PRINTEMPS. The work of Miguel E. Andr\'es has been supported by the LIX-Qualcomm  postdoc fellowship 2010.}
}
\author{ M\'ario S. Alvim  \and Miguel E. Andr\'es,  \\ Konstantinos Chatzikokolakis  \and Catuscia~Palamidessi }
\institute{
 INRIA and LIX, Ecole Polytechnique, France. 
}

\maketitle

\begin{abstract}

Differential privacy is a notion that has emerged in the community of statistical databases, as a response to the problem of protecting the privacy of the database's participants  when performing statistical queries. 
The idea is that a randomized query satisfies differential privacy if the likelihood of obtaining a certain answer for a database $x$ is not too different from the likelihood of obtaining the same answer on adjacent databases, i.e. databases  which differ from $x$ for only one individual. 

Information flow is an area of Security concerned with the problem of controlling the leakage of confidential information in programs and protocols. 
Nowadays, one of the most established approaches  to quantify and  to reason about leakage is based on the R\'enyi min entropy version of  information theory.  

In this paper, we analyze critically the notion of differential privacy in light of the conceptual framework provided by  the R\'enyi min  information theory.
We show that  there is a close relation between differential privacy and leakage, due to the graph symmetries induced by the adjacency relation. 
Furthermore, we consider  the utility of the randomized answer, which measures its expected  degree of accuracy. We  focus on certain kinds of utility functions called ``binary'', which have a 
close correspondence with the R\'enyi min mutual information. Again, it turns out that there can be a tight correspondence between differential privacy and utility, depending 
on the  symmetries induced by the adjacency relation and by the query. Depending on these symmetries we can also build an optimal-utility randomization mechanism while preserving the required level of differential privacy. Our main contribution is a study of the kind of structures that can be induced  by the adjacency relation and the query, and how to use them to derive bounds on the leakage and achieve the optimal utility. 
\end{abstract}

\section{Introduction}\label{section:introduction}
Databases are commonly used for obtaining statistical information about their participants. Simple examples of statistical queries are, for instance, the predominant disease of a certain population,  or the average salary. 
The fact that the answer is publicly available, however, constitutes a threat  for the privacy of the individuals. 

In order to illustrate the problem, consider a set of individuals  $\mathit{Ind}$ whose attribute of interest\footnote{In general we could be interested in several attributes simultaneously, and in this case $\mathit{Val}$ would be a set of tuples.} has values in $\mathit{Val}$. 
A particular database is formed by a subset of $\mathit{Ind}$, where a certain value in $\mathit{Val}$ is associated to each participant. 
A query  is a function $f: {\cal X} \rightarrow {\cal Y}$, where  $\cal X$ is the set of all possible databases, and 
${\cal Y}$ is the domain of the answers. 

For example, let $\mathit{Val}$   be the set of possible salaries and let $f$ represent the query ``what is  the average salary of the participants in the database''. 
In principle we would like to consider the \emph{global information} relative to a database $x$ as \emph{public}, and the \emph{individual information} about a participant $i$ as \emph{private}.  Namely, we would like to be able to obtain $f(x)$ without being able to infer the salary of $i$.  However, this is not always possible. In particular, if the number of participants in $x$ is known (say  $n$), then the removal of $i$  from the database would allow to infer $i$'s salary by querying again the  new database $x'$, and by applying the formula $n\, f(x) -  (n-1)\, f(x')$. Using an analogous reasoning we can argue that not only the removal, but also the addition of an individual is a threat for his privacy. 

Another kind of private information we may want to protect is whether an individual $i$ is participating or not in a database. In this case, if we know for instance that $i$ earns, say  $5K$ Euros/month, and all the other individuals in $\mathit{Ind}$ earn less than $4K$ Euros/month, then knowing that $f(x) > 5K$ Euros/month will reveal immediately that $i$ is in the database $x$. 

A common solution to the above problems is to introduce some output perturbation mechanism based on \emph{randomization}: instead of the exact answer $f(x)$ we report a ``noisy'' answer. Namely, we use some  randomized function ${\cal K}$ which produces values in some domain\footnote{The new domain $\cal Z$ may coincide with $\cal Y$, but not necessarily. It depends on how the randomization mechanism is defined.}  ${\cal Z}$
according to some probability distribution that depends on the input $x\in {\cal X}$. Of course for certain distributions it may still be possible to guess the value of an individual with a high probability of success. 
The notion of \emph{differential privacy}, due to  Dwork \cite{Dwork:06:ICALP,Dwork:09:STOC,Dwork:10:SODA,Dwork:11:CACM}, is a proposal to control  the risk of violating privacy for both kinds of threats described above (value and participation).  
The idea is to say  that $\cal K$ satisfies $\epsilon$-differential privacy  (for some $\epsilon>0$) if the ratio between the probabilities that two adjacent databases give the same answer is bound by $e^\epsilon$, where by ``adjacent'' we mean that the  databases differ for only one individual 
(either for the value of an individual or for the presence/absence of an individual). Often we will abbreviate ``$\epsilon$-differential privacy'' as $\epsilon$-d.p.

Obviously, the smaller is $\epsilon$,  the greater is the privacy protection. In particular, when $\epsilon$ is close to $0$ the output of $\cal K$ is nearly independent from the input (all distributions are almost equal). Unfortunately, such $\cal K$ is practically useless. 
The \emph{utility}, i.e. the capability to retrieve accurate answers from the reported ones, is the other important characteristic of $\cal K$, and it is clear that there is a trade-off between utility and privacy. On the other hand, these two notions are not the complete opposite of each other, because utility concerns the relation between the reported answer and the real answer, while privacy is concerns the relation between the reported answer and the information in the database. This asymmetry makes more interesting the problem of finding a good compromise between the two. 

At this point, we would like to remark an intriguing analogy between the area of differential privacy and that of \emph{quantitative information flow} (QIF), both in the motivations and in the basic conceptual framework. 
Information flow is concerned with the leakage of secret information through computer systems, and the attribute ``quantitative'' refers to the fact that we are interested in measuring the amount of leakage, not just its occurrence. 
One of the most established approaches to QIF is based on information theory: the idea is that 
a system is seen as a channel in the information-theoretic sense, where the secret is the input and the observables are the output. The entropy of the input represents its vulnerability, i.e. how easy it is for an attacher to guess the secret. 
We distinguish between the \emph{ a priori} entropy (before the observable) and the \emph{a posteriori} entropy (given the observable). The difference between the two gives the \emph{mutual information} and represents, intuitively, the increase in vulnerability 
due to the observables produced by the system, so it is naturally considered as a measure of the leakage. 
The notion of entropy is related to the kind of attack we want to model, and in this paper we focus on the  R\'enyi min entropy \cite{Renyi:61:Berkeley}, which represents the so-called \emph{one-try attacks}. 
In recent years there has been a lot of research aimed at establishing the foundations of this framework \cite{Smith:09:FOSSACS,Braun:09:MFPS,Kopf:10:CSF,Andres:10:TACAS,Boreale:11:FOSSACS}.
It is worth pointing out that the a posteriori R\'enyi min entropy corresponds to the concept of Bayes risk, which has also been proposed as a measure of the effectiveness of attacks \cite{Chatzikokolakis:07:CSF,Braun:08:FOSSACS,McIver:10:ICALP}.

The analogy hinted above between differential privacy and QIF is based on the following observations:
at the motivational level, the concern about privacy is akin the concern about information leakage.  
At the conceptual level, the randomized function $\cal K$ can be seen as an information-theoretic channel, 
and the limit case of $\epsilon = 0$, for which the privacy protection is total,  
corresponds to a $0$-capacity channel\footnote{The channel capacity is the maximum mutual information over all possible input distributions.} (the rows of the channel matrix are all identical), which does not allow any leakage. Another promising similarity is that the notion of utility (in the binary case) corresponds closely to the Bayes risk. 

In this paper we investigate the notion of differential privacy, and its implications, in light of the min-entropy information theoretic framework developed for QIF. In particular, we wish to explore  the following natural questions: 
\begin{enumerate}
\item Does $\epsilon$-d.p. induce a bound on the information leakage of $\cal K$? 
\item Does $\epsilon$-d.p. induce a bound on the information leakage \emph{relative to an individual}? 
\item Does $\epsilon$-d.p. induce a bound on the utility? 
\item Given $f$ and $\epsilon$, can we construct a $\cal K$ which satisfies $\epsilon$-d.p. and maximum utility?
\end{enumerate}
We will see that the answers to (1) and (2) are positive, and we provide bounds that are tight, in the sense that for every $\epsilon$ there is a $\cal K$ whose leakage reaches the bound. 
For (3) we are able to give a tight bound in some cases which depend on the structure of the query, and for the same cases, we are able to construct an oblivious\footnote{A randomized function  $\cal K$ is oblivious if its probability distribution depends only on the answer to the query, and not on the database.} $\cal K$ with maximum utility, as requested by (4). 

Part of the above results have already appeared in \cite{Alvim:11:TechRep}, and are based on techniques which exploit the graph structure that the adjacency relation induces on the domain of all databases $\cal X$, and 
on the domain of the correct answers $\cal Y$. The main contribution of this paper is an extension of those techniques, and a coherent graph-theoretic framework for reasoning about the symmetries of those domains. More specifically: 
\begin{itemize}
\item We explore the  graph-theoretic foundations of the adjacency relation, and point out various types of symmetries which allow us to establish a strict link between differential privacy and information leakage. 
\item We give a tight bound for the question (2) above, strictly smaller than the one in \cite{Alvim:11:TechRep}. 
\item We extend the structures for which we give a positive answer to the questions (3) and (4) above. In \cite{Alvim:11:TechRep}
the only case considered was the class  of graphs with single-orbit automorphisms. Here we show that the results hold also for regular-distance graphs and a variant of vertex-transtive graphs. 
\end{itemize}

In this paper we focus on the case in which $\calx$, $\caly$ and $\calz$ are finite, leaving the more general case for future work. 

\section{Preliminaries}
\label{section:preliminaries}

\subsection{Database domain and Differential privacy}

Let $\mathit{Ind}$ be a finite set of individuals that may participate in a database and $\mathit{Val}$ a finite
set of possible values for the attribute of interest of these individuals. In order to capture in a uniform way the presence/absence of an individual in the database, as well as its value, 
we enrich the set of possible values with an element $a$ representing the absence of the individual. Thus the set of all possible databases is the set $\calx = V^\mathit{Ind}$, where $V = \mathit{Val} \cup\{a\}$. 
We will use $u$ and $v$ to denote the cardinalities of  $\mathit{Ind}$ and $\mathit{V}$, $|\mathit{Ind}|$ and $|\mathit{V}|$, respectively. Hence we have that $|\calx| = v^\mathit{u}$. 
A database $x$ can be represented as a 
$u$-tuple $v_0v_1\ldots v_{u-1}$ where each $v_i \in V$ is the value of the corresponding individual. 
Two databases $x,x'$ are \emph{adjacent} (or \emph{neighbors}), written $x\sim x'$, if they differ for the value of exactly one individual. For instance, for $u=3$, $v_0v_1v_2$ and $v_0w_1 v_{2}$, with $w_1 \neq v_1$, are adjacent. 
The structure $(\calx,\sim)$ forms an undirected graph.

Intuitively, differential privacy is based on the idea that a randomized query function  provides sufficient protection if the ratio between the probabilities of two adjacent databases to give a certain answer is bound by $e^\epsilon$, for some given $\epsilon > 0$. 
Formally:
\begin{definition}[\cite{Dwork:11:CACM}]
	\label{def:diff-privacy-1}
	A randomized function $\mathcal{K}$ from $\calx$ to $\calz$ satisfies {$\epsilon$-differential privacy} if for all pairs $x,x'\in \calx$, with $x\sim x'$, and all $S \subseteq \calz$, we have that:
	\begin{equation*}
		\mathit{Pr}[\mathcal{K}(x) \in S] \leq e^{\epsilon} \times \mathit{Pr}[\mathcal{K}(x') \in S]		
	\end{equation*}	
\end{definition}

The above definition takes into account the possibility that $\calz$ is a continuous domain. In our case, since $\calz$ is finite, the probability distribution is discrete, and we can rewrite the property of $\epsilon$-d.p.  more simply as (using the notation of conditional probabilities, and considering both quotients): 
\[\frac{1}{e^\epsilon} \leq  \frac{\mathit{Pr}[Z=z|X=x] }{\mathit{Pr}[Z=z|X=x'] } \leq e^\epsilon  \qquad \mbox{ for all $x,x'\in\calx$ with $x\sim x'$, and all $z\in \calz$}\]
where $X$ and $Z$ represent the random variables associated to $\calx$ and $\calz$, respectively. 

\subsection{Information theory and application to information flow}

In the following, $X, Y$ denote two discrete random variables with carriers
${\cal X} = \{x_{0}{}, \ldots, x_{n-1}{}\}$, ${\cal Y} = \{ y_{0}{}, \ldots, y_{m-1}{} \}$, and
probability distributions $p_{X}(\cdot)$,  $p_{Y}(\cdot)$, respectively. An information-theoretic channel is constituted by an input $X$, an output $Y$, and the matrix of conditional probabilities $p_{Y \mid X}(\cdot \mid \cdot)$, where $p_{Y \mid X}(y \mid x)$ represent the probability that $Y$ is $y$ given that $X$ is $x$. We shall omit the subscripts on the probabilities when they are clear from the context.

 \subsubsection{R\'enyi min-entropy}
	
	In \cite{Renyi:61:Berkeley}, R\'enyi introduced an one-parameter family of entropy measures,  intended as a generalization of Shannon entropy. The R\'enyi entropy of order $\alpha$ ($\alpha > 0$,  $\alpha \neq 1$) of  a random variable $X$ is defined as $H_\alpha(X) \ =\  \frac{1}{1-\alpha}\log_2\sum_{x \,\in\,{\cal X}} p(x)^\alpha$. We are particularly interested in the limit of $H_\alpha$ as $\alpha$ approaches $\infty$. This is  called \emph{min-entropy}. It can be proven  that $H_\infty(X) \ \stackrel{\rm def}{=}\ \lim_{\alpha\rightarrow \infty}H_\alpha(X) \ =\  - \log_2\,\max_{ x\in{\cal X}}\,p(x)$.
	
	R\'enyi defined also the $\alpha$-generalization of other information-theoretic notions, like the Kullback-Leibler divergence. However, he did not define the $\alpha$-generalization of the  conditional entropy, and there is no general agreement on what it should be. For the case $\alpha = \infty$, we adopt here the definition of conditional entropy proposed by Smith in \cite{Smith:09:FOSSACS}:

	\begin{equation}\label{eqn:SmithCondEntropyInfty}
			H_\infty(X\mid Y) \ = \  - \log_2 \sum_{y\in {\cal Y}} p(y)\max_{x\in {\cal X}} \ p(x \mid y)
	\end{equation}
	
	Analogously  to the Shannon case, we can define the R\'enyi-mutual information $I _\infty$ as $H_\infty(X) - H_\infty(X\mid Y)$, and the capacity $C_\infty$ as $\max_{p_{X}(\cdot)}I_\infty(X;Y)$. It has been proven in~\cite{Braun:09:MFPS} that $C_\infty$ is obtained at the uniform distribution, and that it is equal to the sum of the maxima of each column in the channel matrix, i.e., $C_\infty = \sum_{y \,\in\,{\cal Y}} \max_{x \,\in\,{\cal X}} p(y\mid x)$.

\paragraph{Interpretation in terms of attacks:}
R\'enyi min-entropy can be related to a model of  adversary who is allowed to ask exactly one question, which must be of the form ``is $X = x?$'' (one-try attacks). More precisely, $H_\infty(X)$ represents the (logarithm of the inverse of the) probability of success for this kind of attacks and with  the best strategy, which consists, of course, in choosing the $x$ with the maximum probability. 

As for $H_\infty(X\mid Y)$, it represents the inverse of the (expected value of the) probability that the same kind of adversary  succeeds in guessing the value of $X$ \emph{a posteriori}, i.e.  after observing the  result of $Y$. The complement of this probability is also known as \emph{Bayes risk}. Since in general $X$ and $Y$ are correlated, observing $Y$ increases the probability of success. Indeed we can prove formally that $H_\infty(X\mid Y) \leq H_\infty(X)$, with equality if and only if $X$ and $Y$ are independent. $I_\infty(X;Y)$ corresponds to the \emph{ratio} between the probabilities of success a priori and a posteriori, which is a natural notion of leakage. Note that $I_\infty(X;Y)\geq 0$, which seems desirable for a good notion of leakage.

\section{Graph symmetries}
In this section we explore some classes of graphs that
allow us to derive a strict correspondence between $\epsilon$-d.p. 
and the a posteriori entropy of the input. 

Let us first recall some basic notions. Given a graph $G=({\cal V}, \sim)$,  the \emph{distance} $d(v,w)$ between two vertices $v,w\in  \cal V$  is the number of edges in a shortest path connecting them. The \emph{diameter}
of $G$ is the maximum distance between any two vertices in $\cal V$.
The degree of a vertex  is the number of edges incident to it.   $G$ is called \emph{regular} if  every vertex has the same degree. A regular graph with vertices of degree $k$  is called a $k$-regular graph.
An automorphism of $G$   is a permutation $\sigma$ of the vertex set $\calx$, such that for any pair of vertices $x,x'$, if $x\sim x'$, then $\sigma(x)\sim\sigma(x')$. If $\sigma$ is an automorphism, and $v$ a vertex, the orbit of $v$ under $\sigma$ is the set $\{v, \sigma(v), \ldots, \sigma^{k-1}(v)\}$ where $k$ is the smallest positive integer such that $\sigma^k(v) = v$. Clearly, the orbits of the vertices under $\sigma$ define a partition of $\cal V$.

The following two definition introduce the classes of graphs that we are interested in. The first class is well known in literature. 
\begin{definition}
Given a graph $G=({\cal V}, \sim)$, we say that $G$ is   distance-regular if there exist integers $b_i,c_i,i=0,...,d$ such that for any two vertices $v,w$ in $\cal V$  with distance $i=d(v,w)$, there are exactly $c_i$ neighbors of $w$ in $G_{i-1}(x)$  and $b_i$ neighbors of $v$ in $G_{i+1}(x)$, where $G_i(x)$ is the set of vertices $y$ of $G$ with $d(x,y)=i$. 
\end{definition}

\begin{figure}[t]%
		\centering
		\subfigure[{\scriptsize Tetrahedral graph}]{
			\includegraphics[width=0.18\columnwidth]{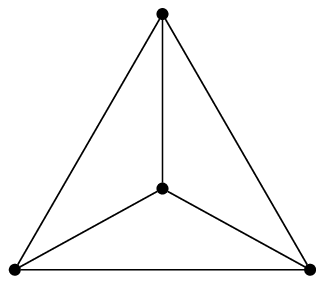}%
			\label{fig:dist-reg3}%
		}
		\qquad
		\subfigure[\scriptsize Cubical graph]{
			\includegraphics[width=0.18\columnwidth]{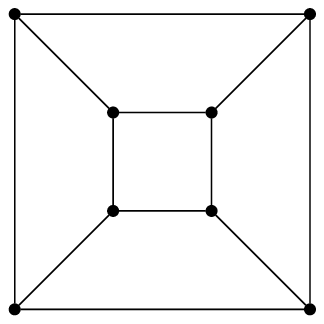}%
			\label{fig:dist-reg4}%
		}
		\qquad
		\subfigure[\scriptsize Petersen graph]{
			\includegraphics[width=0.18\columnwidth]{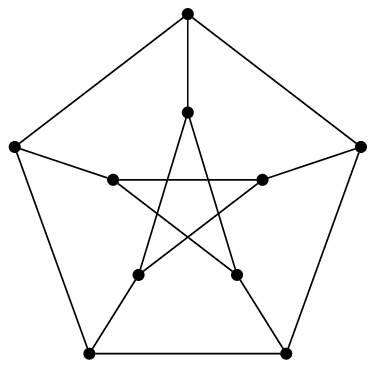}%
			\label{fig:dist-reg5}%
		}	
		\caption{Some distance-regular graphs with degree $3$.}
		\label{fig:dist-reg}	
	\end{figure}

Some examples of distance-regular graphs are illustrated in Figure~\ref{fig:dist-reg}.

The next class is a variant of the VT (vertex-transitive) class: 

\begin{definition}
A graph $G = ({\cal V},\sim)$ is   VT$^+$ (vertex-transitive +) if there are $n$ automorphisms $\sigma_0$, $\sigma_1$, \ldots $\sigma_{n-1}$, where $n = |{\cal V}|$,  such that, for every vertex $v\in {\cal V}$, 
we have that $\{\sigma_i(v)\mid 0\leq i \leq n-1\} = {\cal V}$. 
\end{definition}

\begin{figure}[th]%
		\centering
		\subfigure[\scriptsize Cycle:  degree $2$.]{
			\includegraphics[width=0.20\columnwidth]{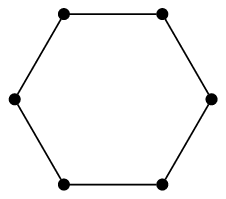}%
			\label{fig:hex1}%
		}
		\qquad
		\subfigure[\scriptsize Degree 4.]{
			\includegraphics[width=0.20\columnwidth]{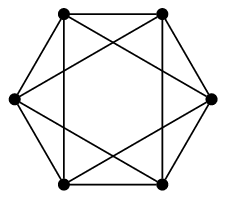}%
			\label{fig:hex2}%
		} 
		\qquad
		\subfigure[\scriptsize Clique: degree 5.]{
			\includegraphics[width=0.20\columnwidth]{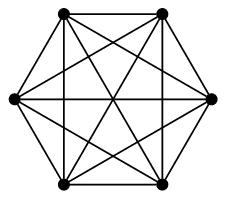}%
			\label{fig:hex3}%
		}	
		\caption{Some VT$^+$ graphs}
		\label{fig:hexagons}	
	\end{figure}

In particular, the graphs  for which there exists an automorphism $\sigma$ which induces only one orbit are VT$^+$: in fact it is sufficient to define $\sigma_i=\sigma^i$ for all $i$ from $0$ to $n-1$. Figure \ref{fig:hexagons} illustrates some graphs with a single-orbit automorphism.

From graph theory we know that neither of the  two classes subsumes the other. They have however a non-empty intersection, which contains in particular  all  the structures of the form $(V^\mathit{Ind},\sim)$, i.e. the database domains.

\begin{proposition}\label{prop:grap-databases}
The structure $({\cal X},\sim) = (V^\mathit{Ind},\sim)$ is both a distance-regular graph  and a VT$^+$ graph.
\end{proposition}

\begin{figure}[th]%
		\centering
		\subfigure[\scriptsize $|\mathit{Ind}|=4, V=\{a,b\}$ ($4$-dimensional  hypercube)]{
			\includegraphics[width=0.37\columnwidth]{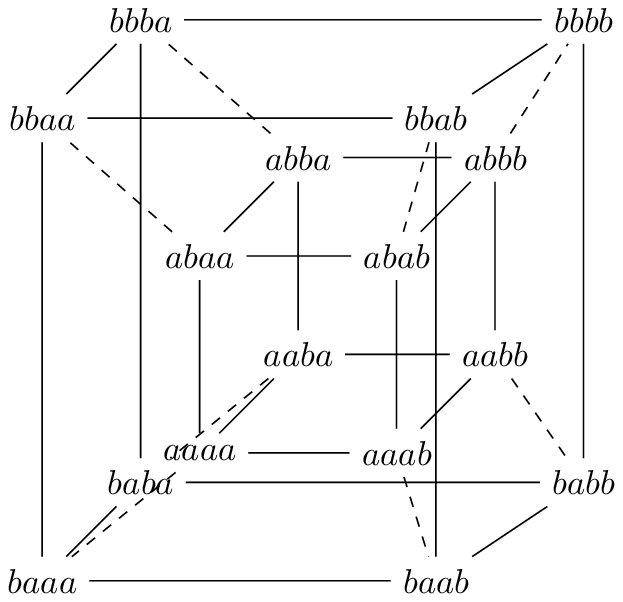}%
			\label{fig:hyp1}%
		}
		\qquad
		\subfigure[\scriptsize $|\mathit{Ind}|=3, V = \{a,b,c\}$ (for readability sake  we show only part of the graph)]{
			\includegraphics[width=0.37\columnwidth]{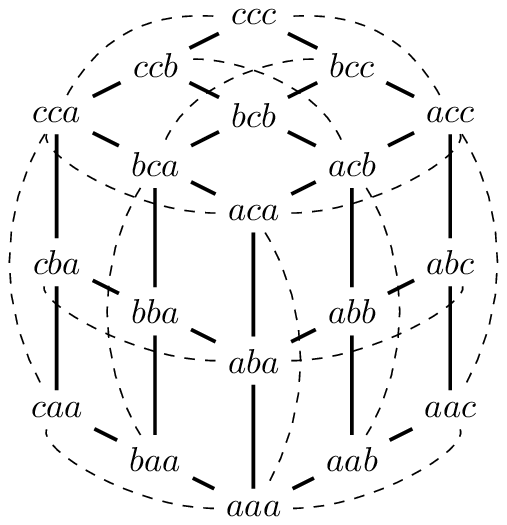}%
			\label{fig:hyp2}%
			}
		\caption{Some $(V^\mathit{Ind},\sim)$ graphs}
		\label{fig:hypercubes}	
		\vspace{.5cm}
	\end{figure}

Figure \ref{fig:hypercubes} illustrates some examples of structures  $(V^\mathit{Ind},\sim)$. Note that when $|\mathit{Ind}| = n$ and $|V|=2$, $(V^\mathit{Ind},\sim)$ is the $n$-dimentional hypercube. 

\begin{figure}[t]%
		\centering
		\includegraphics[width=0.4\columnwidth]{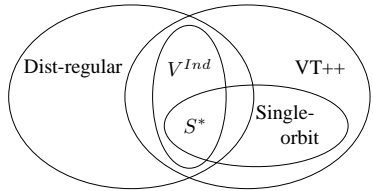}%
		\caption{Venn diagram for the classes of graphs considered in this section. Here, $S^* = \{V^{Ind} \ | \ |V| =  2, |Ind| \leq 2 \}$}%
		\label{fig:venn}%
	\end{figure}

The situation is summarized in Figure \ref{fig:venn}. We remark that in general the graphs $(V^\mathit{Ind},\sim)$ do not have a single-orbit automorphism. The only exceptions are the two simplest structures ($|V|=2, |\mathit{Ind}| \leq 2$).

The  two symmetry classes defined above, distance-regular and VT$^+$,  will be used in the next section to transform a generic channel matrix into a matrix with a symmetric structure, while preserving the a posteriori min entropy and the $\epsilon$-d.p.. This is the core of our technique to establish the relation between differential privacy and quantitive information flow, depending on the structure induced by the database adjacency relation.

\section{Deriving the relation between differential privacy and QIF on the basis of the graph structure}

This section contains  the main technical contribution of the paper: a general technique for determining the relation  between $\epsilon$-differential privacy and leakage, and between  $\epsilon$-differential privacy and utility, depending on the graph structure induced by $\sim$ and $f$. 
The idea is to use the symmetries of the graph structure to  transform the channel matrix into an equivalent  matrix with certain regularities, which allow to establish the link between $\epsilon$-differential privacy and  the a posteriori min entropy. 

\begin{figure}[t]%
		\centering
		\includegraphics[width=0.5\columnwidth]{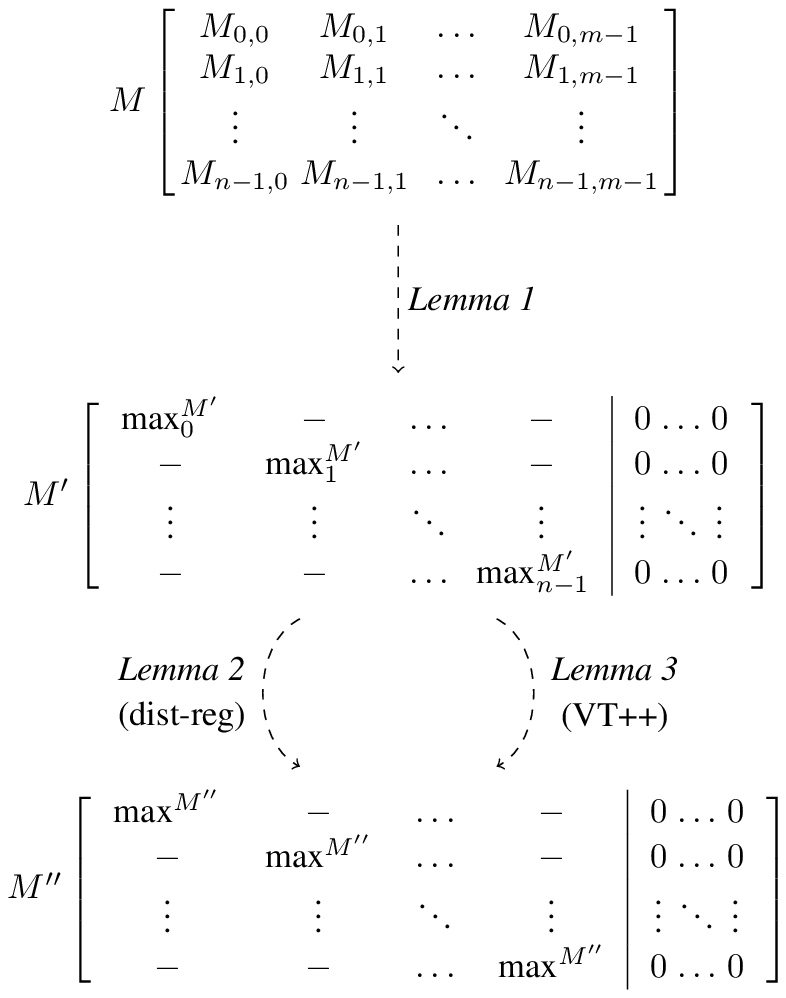}%
		\caption{Matrix transformations for distance-regular and VT$^+$ graphs}%
		\label{fig:mat-transf}%
	\end{figure}

Let us illustrate briefly this transformation. Consider a channel whose matrix $M$ has at least as many columns as rows.   First, we transform $M$ into a matrix $M'$ in which each of the first $n$ columns has a maximum  in the diagonal, and the remaining columns are all $0$'s. 
Second, under the assumption that the input domain  is distance-regular or VT$^+$, we transform $M'$ into a matrix $M''$ whose diagonal elements are all the same, and coincide with the maximum element of $M''$, which we  denote here by $\text{max}^{M''}$. 
These steps are illustrated in Figure \ref{fig:mat-transf}.

We are now going to present formally our the technique. Let us first fix some notation: In the rest of this section  we consider channels with input  $A$ and output  $B$, with carriers $\cal A$ and $\cal B$ respectively, and we assume that the probability distribution of $A$ is uniform. Furthermore, we assume that $|{\cal A}| = n \leq |{\cal B}| = m$.
We also assume  an adjacency relation $\sim$ on  $\cal A$, i.e.  that $({\cal A}, \sim)$ is an undirected  graph structure. With a slight abuse of notation, we will also write $i\sim h$ when $i$ and $h$ are associated to adjacent elements of $\cal A$, and we will write $d(i,h)$ to denote the distance between the elements of $\cal A$ associated to $i$ and $h$. 

We note that a channel matrix $M$ satisfies $\epsilon$-d.p.  if for each column $j$ and for each pair of rows $i$ and $h$ such that $i\sim h$ we have that:
\[
\frac{1}{e^\epsilon}\leq \frac{M_{i,j}}{M_{h,j}}\leq e^\epsilon.
\]
The a posteriori entropy of a channel with matrix $M$ will be denoted by $H^M_\infty(A|B)$. 

Next Lemma is relative to the first step of the transformation. 
\begin{lemma}\label{lem:transform-diagonal}
Consider a channel with matrix $M$. Assume that $M$ satisfies $\epsilon$-d.p.. Then it is possible to transform $M$ into a matrix $M'$ such that: 
\begin{itemize}
\item Each of the first $n$ columns has a maximum in the diagonal, i.e. $M'_{i,i}=  \max^{M'}_{i} = \max_h M'_{h,i}$ for each $i$ from $0$ to $n-1$. 
\item The rest of the columns contain only $0$'s, i.e. $M'_{i,j}= 0$ for each $i$ from $0$ to $n-1$ and each $j$ from $n$ to $m-1$.
\item $M'$ satisfies $\epsilon$-d.p.
\item $H^{M'}_\infty(A|B) = H^{M}_\infty(A|B)$. 
\end{itemize}
\end{lemma}

Next lemma is relative to the second step of the transformation, for the case of distance-regular graphs. 

\begin{lemma}\label{lem:transform-dr}
Consider a channel with matrix $M'$. Assume that $M'$ satisfies $\epsilon$-d.p., and the first $n$ columns have  maxima   in the diagonal,  and the rest of the columns contain only $0$'s. Assume that $({\cal A}, \sim)$ is distance-regular. Then it is possible to transform $M'$ into a matrix $M''$ such that: 
\begin{itemize}
\item The elements of the diagonal are all the same, and are equal to the maximum of the matrix, i.e. $M''_{i,i}= \max^{M''}=\max_{h,i} M''_{h,i}$ for each $i$ from $0$ to $n-1$. 
\item The rest of the columns contain only $0$'s.
\item $M''$ satisfies $\epsilon$-d.p.
\item $H^{M''}_\infty(A|B) = H^{M'}_\infty(A|B)$. 
\end{itemize}
\end{lemma}

Next lemma is relative to the second step of the transformation, for the case of VT$^+$ graphs. 
\begin{lemma}\label{lem:transform-vt++}
Consider a channel with matrix $M'$ satisfying the assumptions of Lemma \ref{lem:transform-dr}, except for the assumption about distance-regularity, which we replace by the assumption that $({\cal A}, \sim)$ is VT$^+$. 
Then it is possible to transform $M'$ into a matrix $M''$ with the same properties as in Lemma~\ref{lem:transform-dr}.
\end{lemma}
Note that the fact that in $M''$ the diagonal elements are all equal to the maximum $ \max^{M''}$ implies that $H^{M''}_\infty(A|B) = \max^{M''}$. 

Once we have a matrix with the properties of $M''$, we can use again the graph structure of $\cal A$ to determine a bound on $H^{M''}_\infty(A|B)$. 

First we note that the property of $\epsilon$-d.p. induces a relation between the ratio of elements at any distance: 

\begin{remark}
Let $M$ be a matrix satisfying  $\epsilon$-d.p.. Then, for any column $j$, and any pair of rows $i$ and $h$ we have that:
\[
\frac{1}{e^{\epsilon \,d(i,h)}}\leq \frac{M_{i,j}}{M_{h,j}}\leq e^{\epsilon \,d(i,h)}
\]
\end{remark}
In particular, if we know that the diagonal elements of $M$ are equal to the maximum element $\max^{M}$, then 
for each element $M_{i,j}$ we have that:
\begin{equation}\label{eqn:other-elements}
M_{i,j}\geq \frac{\max^{M}}{\displaystyle e^{\epsilon \,d(i,j)}}
\end{equation}
Let us fix a row, say row $r$. For each distance $d$ from $0$ to the diameter of the graph, let $n_d$ be the number of elements $M_{r,j}$ that are at distance $d$ from the corresponding diagonal element $M_{j,j}$, i.e. such that $d(r,j) = d$. (Clearly, $n_d$ depends on the structure of the graph.) Since the elements of the row $i$ represent a probability distribution, we obtain the following dis-equation: 

\[  {\textstyle \max^{M}} \sum_{d}  \frac{n_d}{e^{\epsilon \,d }} \leq 1\]
from which we derive immediately a bound on the min a-posteriori entropy. 

Putting together all the steps  of this section, we  obtain our main result. 

\begin{theorem}\label{theo:bound}
Consider a matrix $M$, and let $r$ be a row of $M$. Assume that $({\cal A}, \sim)$ is either  distance-regular or VT$^+$, and that $M$ satisfies $\epsilon$-d.p.
For each distance $d$  from $0$ to the diameter of $({\cal A}, \sim)$,  let $n_d$ be the number of nodes $j$ at distance $d$ from $r$. 
Then we have that:
\begin{equation}\label{eqn:bound}
H^{M}_\infty(A|B) \geq - \log_2 \frac{1}{\displaystyle  \sum_{d}  \frac{n_d}{e^{\epsilon \,d }}} 
\end{equation}
\end{theorem}

Note that this bound is tight, in the sense that we can build a matrix for which (\ref{eqn:bound}) holds with equality. 
It is sufficient to define each element $M_{i,j}$ according to (\ref{eqn:other-elements}) (with equality instead of dis-equality, of course). 

In the next section, we will see how to use this theorem for establishing a bound on the leakage and on the utility. 

\section{Application  to leakage}
\label{section:leakage}
As already hinted in the introduction, we can regard $\cal K$ as a channel with input $X$ and output $Z$. 
From Proposition \ref{prop:grap-databases} we know that  $(\calx, \sim)$ is both distance-regular and VT$^+$, we can therefore apply Theorem~\ref{theo:bound}. 
Let us fix a particular  database $x\in \calx$. The number of databases at distance $d$ from $x$ is 
\begin{equation}\label{eqn:coeff} 
n_d =  \left(\begin{array}{c}u\\d\end{array}\right)\,(v-1)^{d}
\end{equation}
where $u=|\mathit{Ind}|$ and $v=\mathit{V}$. In fact, recall that $x$ 
can be represented as a $u$-tuple with values in $V$. 
We need to select $d$ individuals in the $u$-tuple and then change their values, and each of them can be changed  in $v-1$ different ways.  

Using the $n_d$ from (\ref{eqn:coeff}) in Theorem~\ref{theo:bound} we obtain a binomial expansion in the denominator, namely:
\[
H^{M}_\infty(X|Z) \geq - \log_2 \frac{1}{\displaystyle \sum_{d=0} ^u  \left(\begin{array}{c}u\\d\end{array} \right) \, (v-1)^{d} \,\frac{e^{\epsilon (u-d)}}{e^{\epsilon \, u}}} = 
- u \,\log_2\frac{e^{\epsilon}}{v-1+e^\epsilon}
\]
which gives the following result: 

\begin{theorem}
If $\cal K$ satisfies $\epsilon$-d.p., then for the uniform input distribution the information leakage is bound from above as follows: 
\[
I_\infty(X;Z)\leq  u\, \log_2\frac{v\,e^{\epsilon}}{v-1+e^\epsilon}
\]
\end{theorem}

We consider now the \emph{leakage for a single individual}. Let us fix a database $x$, and a particular individual $i$ in $\mathit{Ind}$. The  possible ways in which we can change the value of $i$ in 
$x$ are $v-1$. All the new databases obtained in this way are adjacent to each other, i.e. the graph structure associated to the input is a clique of $v$ nodes. 
Therefore we obtain $n_d=1$ for $d=0$, $n_d= v-1$ for $d=1$, and $n_d=0$ otherwise. 
By substituting this value of $n_d$ in Theorem~\ref{theo:bound}, we get
\[
H^{ind}_\infty(\mathit{Val}|Z) \geq - \log_2 \frac{1}{\displaystyle  1 +   \frac{v-1}{e^{\epsilon}}}  = - \log_2\frac{e^\epsilon}{v-1 + e^\epsilon}
\]
which leads to the following result:
\begin{proposition}\label{prop:ind}
Assume that $\cal K$ satisfies $\epsilon$-d.p.. Then for the uniform distribution on $\mathit V$ the information leakage for an individual is bound from above as  follows: 
\[
I^{ind}_\infty(\mathit{Val};B)\leq  \log_2\frac{v\,e^{\epsilon}}{v-1+e^\epsilon}
\]
\end{proposition}
Note  that the bound on the leakage for an individual does not depend on the size of $\mathit{Ind}$, nor on the database $x$ that we fix.
		
\section{Application to utility}
\label{sec:utility}
We turn now our attention to the issue of \emph{utility}. We focus on the case in which $\cal K$ is \emph{oblivious}, which means that it depends only on the (exact) answer to the query, i.e. on the value of $f(x)$, and not on $x$.  
 
\begin{figure}[t]%
		\centering
		\includegraphics[width=.8\columnwidth]{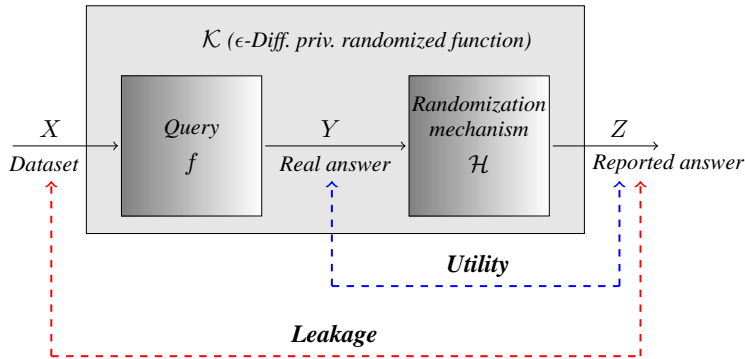}%
		\caption{Schema of an oblivious randomized function}%
		\label{fig:ut-leak-scheme}%
	\end{figure}

An oblivious  function can be decomposed in the concatenation of two channels, one representing the function $f$, and the other representing the randomization mechanism $\cal H$ 
added  as output perturbation. The situation is illustrated in Figure~\ref{fig:ut-leak-scheme}. 

The standard way to define utility  is by means of $\mathit{guess}$ and $\mathit{gain}$ functions. The functionality of the first is 
$\mathit{guess}:\calz\rightarrow \caly$, and it represents the user's strategy to retrieve the correct answer form the reported one. The functionality of the latter is $\mathit{gain}:\caly\times\caly\rightarrow \mathbb{R}$. the value $\mathit{gain}(y,y')$ represents  the reward for guessing the answer $y$ when the correct answer is $y'$. 
The utility $\calu$ can then be defined as the expected gain: 
\[
\calu(Y,Z) = \sum_{y,z} p(y,z)\,\mathit{gain}(\mathit{guess}(z),y)
\]
We focus here on the so-called  \emph{binary} gain function, which is defined as
\[
\mathit{gain}(y,y') = \left\{\begin{array}{lll} 
					1 && \mbox{if } y = y'\\[1mm]
					0 && \mbox{otherwise }
				\end{array}
				\right.
\]

This kind of function represents the case in which there is no reason to prefer an answer over the other, except if it is the \emph{right} answer. More precisely,  we get a gain if and only if we guess the right answer. 

If the gain function is binary, and the $\mathit{guess}$ function represents the user's best strategy, i.e. it is chosen to optimize  utility, then there is a well-known correspondence between  $\calu$ and the Bayes risk / the a posteriori min entropy. Such correspondence   is  expressed by the following proposition:

\begin{proposition}
Assume that $\mathit{gain}$ is binary and  $\mathit {guess}$ is optimal. Then:
\[\calu(Y,Z) = \sum_{z} \max_y (p(z|y) \, p(y)) = 2^{-H_\infty(Y|Z)}\]
\end{proposition}

In order to analyze the implications of the  $\epsilon$-d.p. requirement on the utility, we need to consider the structure that the adjacency relation induces on $\caly$. Let us define $\sim$ on $\caly$ as follows: $y\sim y'$ if there are $x,x' \in\calx$ such that $y = f(x)$, $y' = f(x')$, and $x\sim x,$. Note that $\cal K$ satisfies $\epsilon$-d.p. 
if and only if $\cal H$ satisfies $\epsilon$-d.p. 

If $(\caly,\sim)$ is distance-regular  or  VT$^+$, then we can apply Theorem~\ref{theo:bound} to find a bound on the utility. In the following, we assume that the distribution of $Y$ is uniform.
\begin{theorem}\label{theo:util}
Consider a randomized mechanism  $\cal H$, and let $y$ be an element    of $\caly$. Assume that $({\caly}, \sim)$ is either  distance-regular or VT$^+$ and that $\cal H$ satisfies $\epsilon$-d.p.
For each distance $d$  from $0$ to the diameter of $({\caly}, \sim)$,  let $n_d$ be the number of nodes $y'$ at distance $d$ from $y$. 
Then we have that:
\begin{equation}\label{eqn:util}
 {\cal U}(Y,Z) \leq   \frac{1}{\displaystyle  \sum_{d}  \frac{n_d}{e^{\epsilon \,d}}} 
\end{equation}
\end{theorem}

The above bound is tight, in the sense that (provided $(\caly,\sim)$ is distance-regular  or  VT$^+$) we can construct a mechanism $\cal H$ which satisfies (\ref{eqn:util}) with equality. 
More precisely, define  
\[c = \frac{1}{\displaystyle  \sum_{d}  \frac{n_d}{e^{\epsilon \,d}}}\]
Then define $\cal H$ (here identified with its channel matrix for simplicity) as follows: 
 \begin{equation}\label{eqn:other-elements-util}
{\cal H}_{i,j}= \frac{c}{\displaystyle e^{\epsilon \,d(i,j)}}
\end{equation}

\begin{theorem}\label{theo:maxutil}
Assume $({\caly}, \sim)$ is   distance-regular or VT$^+$. Then the matrix    $\cal H$ defined in (\ref{eqn:other-elements-util}) satisfies $\epsilon$-d.p.
and has maximal utility: 
\begin{equation*}\label{eqn:maxutil}
 {\cal U}(Y,Z) =   \frac{1}{\displaystyle  \sum_{d}  \frac{n_d}{e^{\epsilon \,d}}} 
\end{equation*}
\end{theorem}

Note that we can always define $\cal H$ as in (\ref{eqn:other-elements-util}): the matrix so defined will be a legal channel matrix, and it will satisfy $\epsilon$-d.p.. 
However, if $({\caly}, \sim)$ is  neither  distance-regular nor VT$^+$, then the utility of such $\cal H$ is not necessarily optimal.

We end this section with an example (borrowed from \cite{Alvim:11:TechRep}) to illustrate our technique. 

\begin{example} Consider a database with electoral information where each row corresponds to a voter and contains the following  three fields:

\begin{itemize}
\item \emph{Id}: a unique (anonymized) identifier assigned to each voter;
\item \emph{City}: the name of the city where the user voted;
\item \emph{Candidate}: the name of the candidate the user voted for.
\end{itemize}

Consider the query \emph{``What is the city with the greatest number of votes for a given candidate $\mathit{cand}$?''}. For such a query the binary  utility function is the natural choice: only the right city gives some gain, and all  wrong answers are equally bad. It is easy to see that every two answers are neighbors, i.e. the graph structure of the answers is a clique. 

Let us consider the scenario where \emph{City} $=\{A,B,C,D,E,F\}$ and assume for simplicity that there is a unique answer for the query, i.e., there are no two cities with exactly the same number of individuals voting for candidate $\mathit{cand}$. Table~\ref{tab:city-cand-geo} shows two alternative mechanisms providing $\epsilon$-differential privacy (with $\epsilon = \log 2$). The first one, $M_{1}$, is based on the truncated geometric mechanism method used in \cite{Ghosh:09:STC} for counting queries (here extended to the case where every pair of answers is neighbor). The second mechanism, $M_{2}$, is obtained by applying the definition  (\ref{eqn:other-elements-util}). From 
Theorem~\ref{theo:maxutil} we know that for the uniform input distribution $M_{2}$ gives optimal utility.

For the uniform input distribution, it is easy to see that ${\cal U}(M_1) = 0.2242 < 0.2857 = {\cal U}(M_2)$. Even for non-uniform distributions, our mechanism still provides better utility. For instance, for $p(A) = p(F) = 1/10$ and $p(B) = p(C) = p(D) = P(E) = 1/5$, we have ${\cal U}(M_1) = 0.2412 < 0.2857 = {\cal U}(M_2)$.
This is not too surprising: the geometric mechanism, as well as the Laplacian mechanism proposed by Dwork, perform  very well when the domain of answers is provided with a metric and the utility function is not binary\footnote{In the metric case  the gain function can take into account the proximity of the reported answer to the real one, the idea being that a close answer, even if  wrong, is better than a distant one.}. It also works well when $({\cal Y}, \sim)$ has low connectivity, in particular in the cases of a ring and of a line. But in this example, we are not in these cases, because we are considering \emph{binary gain functions} and \emph{high connectivity}. 

\begin{table}[tb]
\centering	
	\subtable[$M_{1}$: truncated geometric mechanism]{	
		$
			\begin{array}{|c||c|c|c|c|c|c|}
				\hline
				\mbox{In/Out} & A    & B    & C    & D    & E    & F    \\ \hline \hline
				A & 0.535 & 0.060 & 0.052 & 0.046 & 0.040 & 0.267 \\ \hline
				B & 0.465 & 0.069 & 0.060 & 0.053 & 0.046 & 0.307 \\ \hline
				C & 0.405 & 0.060 & 0.069 & 0.060 & 0.053 & 0.353 \\ \hline
				D & 0.353 & 0.053 & 0.060 & 0.069 & 0.060 & 0.405 \\ \hline
				E & 0.307 & 0.046 & 0.053 & 0.060 & 0.069 & 0.465 \\ \hline
				F & 0.267 & 0.040 & 0.046 & 0.052 & 0.060 & 0.535 \\ \hline
			\end{array}
		$
		\label{tab:city-cand-geo-a}
	}	\ \ \ \ \ \ 
	\subtable[$M_{2}$: our mechanism]{		
		$
			\begin{array}{|c||c|c|c|c|c|c|}
			 \hline
			 \mbox{In/Out} & A    & B    & C    & D    & E    & F    \\ \hline \hline
			 A             & 2/7  & 1/7 & 1/7 & 1/7 & 1/7 & 1/7 \\ \hline
			 B             & 1/7  & 2/7 & 1/7  & 1/7 & 1/7 & 1/7 \\ \hline
			 C             & 1/7 & 1/7 & 2/7  & 1/7  & 1/7 & 1/7 \\ \hline
			 D             & 1/7  & 1/7 & 1/7  & 2/7  & 1/7  & 1/7  \\ \hline
			 E             & 1/7 & 1/7 & 1/7 & 1/7  & 2/7 & 1/7 \\ \hline
			 F             & 1/7 & 1/7 & 1/7 & 1/7 & 1/7 & 2/7  \\ \hline
			\end{array}
		$
		\label{tab:city-cand-geo-b}
	}
	\caption{Mechanisms for the city with higher number of votes for candidate $\mathit{cand}$}
	\label{tab:city-cand-geo}
\end{table}

\end{example}

\section{Related work}
\label{section:related-work}
As far as we know, the first work to investigate the relation between
differential privacy and information-theoretic leakage \emph{for an
individual} was \cite{Alvim:10:TechRep}. In this work, a channel is relative to a given
database $x$, and the channel inputs  are all possible databases
adjacent to $x$. Two bounds on leakage were presented, one for teh R\'enyi min entropy, and one for Shannon entropy. 
Our bound in Proposition~\ref{prop:ind} is an improvement
with respect to the (R\'enyi min entropy) bound in \cite{Alvim:10:TechRep}.

Barthe and K\"opf \cite{Barthe:11:CSF} were the first to investigates the (more
challenging) connection between differential privacy and the R\'enyi
min-entropy leakage \emph{for the entire universe of possible databases}.
They consider the ``end-to-end differentially private mechanisms'',
which correspond to what we call $\cal K$ in
our paper, and propose, like we do, to interpret them as information-theoretic
channels. 
They provide a bound for the leakage, but  point out that it  is not tight in general, and show that there cannot be a
domain-independent bound, by proving that for any number of individual
$u$ the optimal bound must be at least a certain expression $f(u,\epsilon)$.
Finally, they show that the question of providing optimal upper bounds
for the leakage of $\epsilon$-differentially private randomized functions in
terms of rational functions of $\epsilon$ is decidable, and leave the actual
function as an open question. In our work we used rather different
techniques and found (independently) the same function $f(u,\epsilon)$ 
(the bound in Theorem~\ref{theo:bound}), but 
 we actually proved  that $f(u,\epsilon)$ is the
optimal bound\footnote{When discussing our result with Barthe and
K\"opf, they said that they also conjectured that $f(u,\epsilon)$ is the
optimal bound.}.
Another difference is that \cite{Barthe:11:CSF}  captures the case in which the focus
of differential privacy is on hiding \emph{participation} of
individuals in a database. In our work, we consider both the participation and    the \emph{values} of the participants.

Clarkson and Schneider also considered differential privacy as a case study of their proposal for quantification of integrity  \cite{Clarkson:11:TECHREP}. There, the authors analyze database privacy conditions from the literature (such as differential privacy, $k$-anonymity, and $l$-diversity) using their framework for utility quantification. In particular, they study the relationship between differential privacy and a notion of leakage (which is different from ours - in particular their definition is based on Shannon entropy) and they provide a tight bound on leakage. 

Heusser and Malacaria \cite{Heusser:09:FAST} were among the first to explore the application of information-theoretic concepts to databases queries. They proposed to model database queries as programs, which allows for statical analysis of the information leaked by the query.  However  \cite{Heusser:09:FAST}  did not attempt to relate information leakage to differential privacy.

In \cite{Ghosh:09:STC} the authors aim at obtaining optimal-utility randomization mechanisms while preserving differential privacy. The authors propose adding noise to the output of the query according to the geometric mechanism. Their framework is very interesting in the sense it provides a general definition of utility for a mechanism $M$ that captures any possible side information and preference (defined as a loss function) the users of $M$ may have. They prove that the geometric mechanism is optimal in the particular case of counting queries. Our results in Section \ref{sec:utility} do not restrict to counting queries, but on the other hand we only consider the case of binary loss function.

\section{Conclusion and future work}
In this paper we have investigated the relation between $\epsilon$-differential privacy and leakage, and between $\epsilon$-differential privacy and utility.
 Our main contribution is the development of a general technique for determining these relations  depending on the graph structure induced by the adjacency relation and by the query.
 We have considered two particular structures, the distance-regular graphs, and the VT$^+$ graphs, which allow to obtain tight bounds on the leakage and on the utility, and to construct the optimal randomization mechanism satisfying $\epsilon$-differential privacy.

As future work, we plan to extend our result to other kinds of utility functions. In particular, we are interested in the case in which the the answer domain is provided with a metric, and we are interested in taking into account  the degree of accuracy of the inferred answer.  

\bibliographystyle{plain}
\bibliography{short}



\end{document}